# Ipotesi astronomiche sul foro della colonna augustea di Santa Maria in Aracoeli

**Costantino Sigismondi**

ICRA-Sapienza Università di Roma, UPRA Istituto Scienza e Fede, and Observatorio Nacional Rio de Janeiro
sigismondi@icra.it

**Abstract:**

The third column of the left row in the Basilica of Santa Maria in Aracoeli in Rome, has a hole carved in it, of probable astronomical use.
The tube aims at a point a 116 degrees of azimut, East-South-East and 17.3 degrees of altitude, the Sun could be aimed in that direction at 07:00 UTC of 12 october and 7:25 UT of march 2, but the view is obstructed by the church's building. Same situation for the 3.9 magnitude alpha Monocerotis, visible 2000 years ago, taking into account the precession.
Other hypotheses on the meridian astronomical use of this hole are here reviewed.
In the case of solar observations it cannot be for meridian transits in Rome, since the lower transits occurs now at 24.5 degrees at winter solstice; we investigate the case of stellar transits, of declination 30.8 degrees South. Many alignments are possible leaving the azimut as free parameter, being the columun not in its original collocation, but here we investigate only the meridian transits, and no bright stars could have been seen through that hole, taking also into account the precession 2000 years ago. We conclude that this hole was probably not used for meridian observations.

**Introduzione:**

La terza colonna della nave sinistra della Basilica di Santa Maria in Aracoeli sul Campidoglio, di porfido, ha un foro non diametrale, inclinato 17.3° sull'orizzontale, ad 1m 61 dal pavimento ed orientato verso l'azimut 116°, Est Sud-Est.
L'angolo di altezza esclude un strumento per l'osservazione del transito meridiano del Sole da Roma, in quanto l'altezza mirata è inferiore a quella del transito al solstizio invernale. Visto che la colonna era originariamente altrove l'orientamento del foro puo' essere diverso dall'attuale e poteva anche indicare un'ora antimeridiana o pomeridiana, di un giorno particolare, in cui il Sole raggiunge quella particolare altezza e azimut: potrebbe indicare un evento come l'ora della nascita di una persona importante... un oroscopo in senso letterale del termine.
Oppure il tubo potrebbe essere stato usato per mirare il transito meridiano di una stella australe, di declinazione δ=-30°48'±70' al tempo di Augusto, o ancora prima.
Si presentano i rilevamenti metrici e angolari di questo tubo scavato nel porfido, con le possibili ipotesi sull'uso astronomico.

**La colonna con il foro**

La colonna in questione porta in alto una scritta A CUBICVLO AUGUSTORUM che ricorda la tradizione che Augusto abbia avuto l'apparizione della Vergine col Bambino nella sua camera da letto.[1]

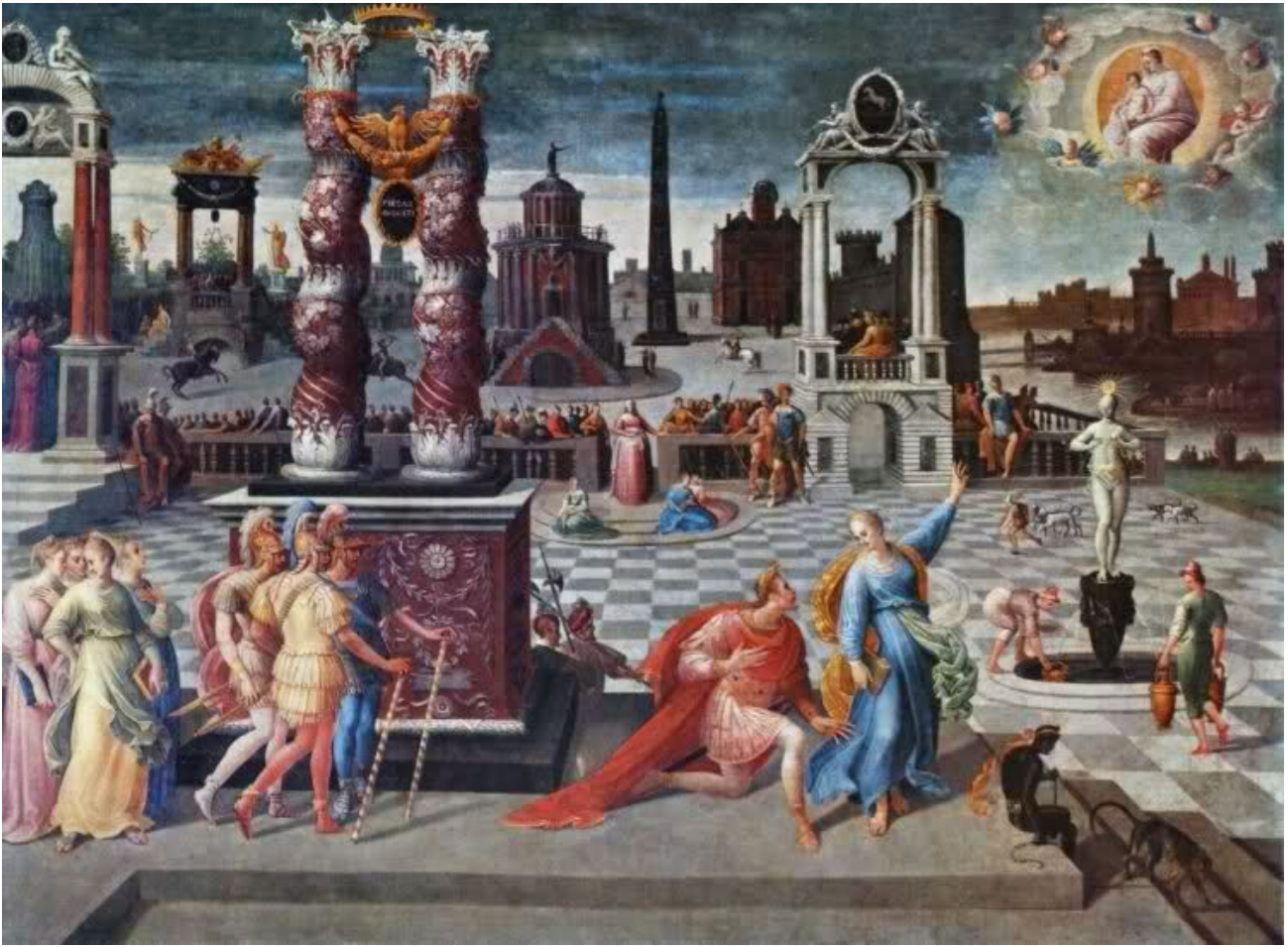

Fig. 1 Rappresentazione della visione di Augusto con l'intermediazione della Sibilla Tiburtina. La visione di Augusto avvenne nella sua camera da letto e la tradizione riporta che l'imperatore fece erigere un altare chiamato altare del cielo, «Ara Coeli»; dipinto di Antoine Caron, 1580, Louvre, Parigi.

---

[1] http://www.ilpatrimonioartistico.it/la-basilica-di-santa-maria-in-aracoeli-2/
http://romaleggendaria.blogspot.it/2009/10/la-leggenda-di-santa-maria-in-aracoeli.html

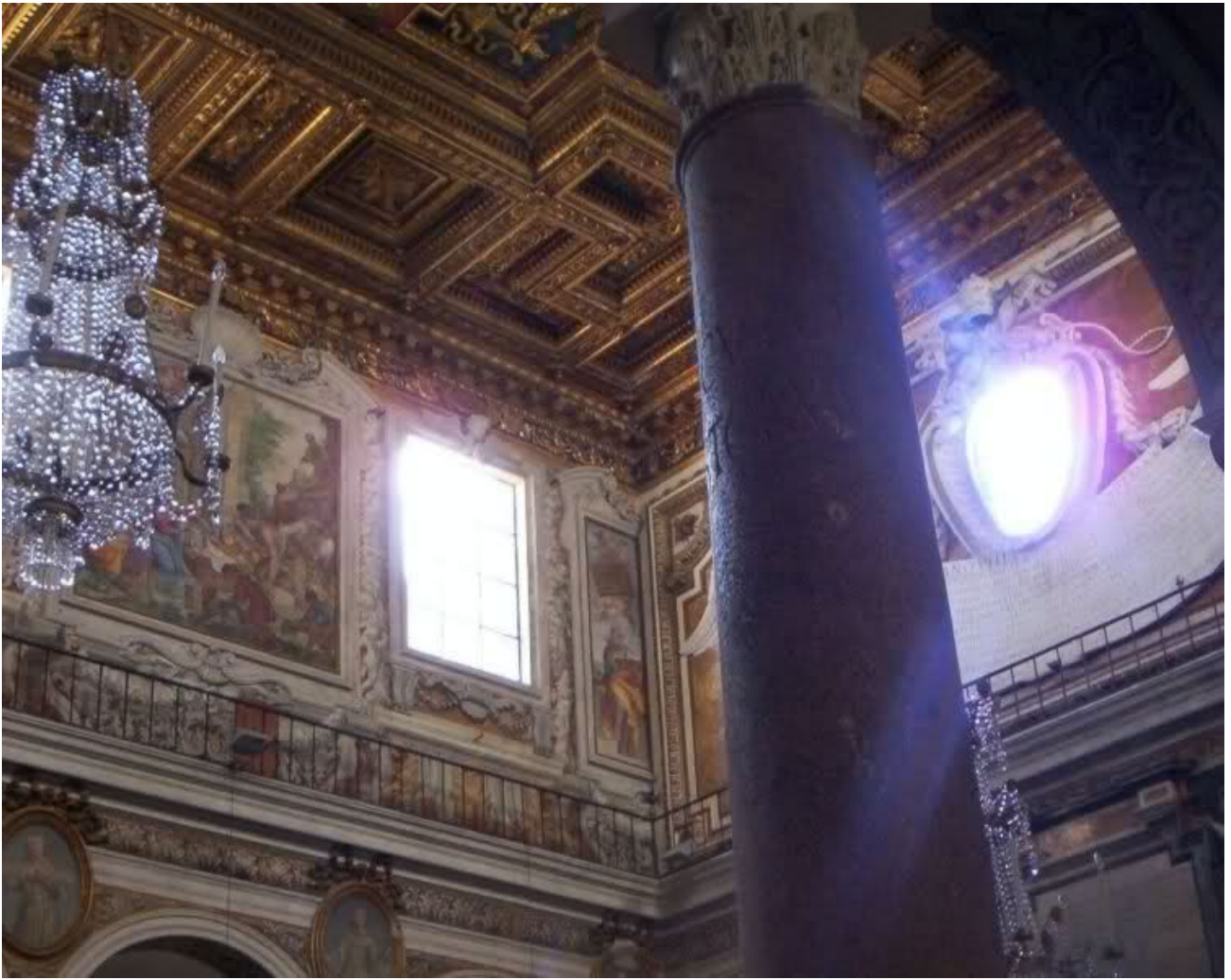

Fig. 2 La colonna in porfido, terza dall'ingresso della chiesa, navata sinistra, con il foro di probabile uso astronomico. In alto la scritta A CUBICVLO AVGUSTORVM ricorda la tradizione del luogo dell'apparizione della Vergine col Bambino ad Augusto, nella sua camera da letto (cubiculum).

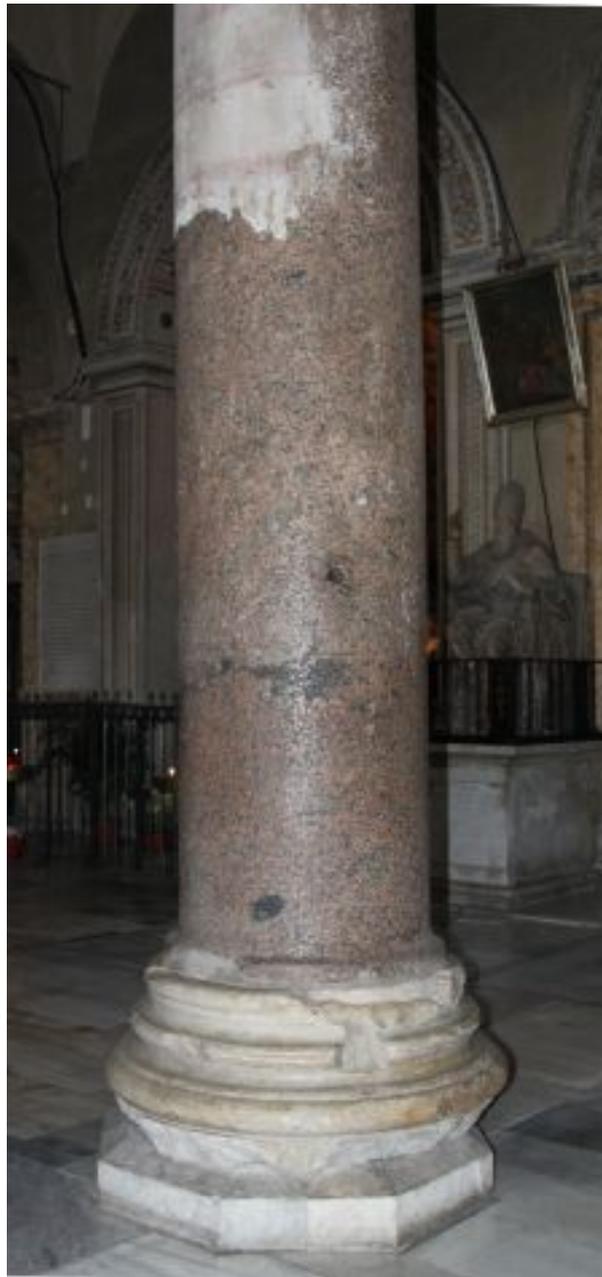

Fig. 3 La colonna con la parte superiore del foro ad 1m 83 dal pavimento.

**La chiesa: orientamento dell'edificio e del foro**

La Chiesa di Santa Maria in Aracoeli si trova sulla sommità settentrionale del colle Capitolino in Roma, sul luogo dove sorgeva il tempio di Giunone Moneta (Ammonitrice) dove era anche la zecca di Roma, da cui ancora oggi abbiamo il termine «moneta». Come spesso accade a Roma gli edifici di culto cristiani conservano orientamento e fondamenta dei precedenti edifici pagani.

Alle 11:30 del 18 gennaio, giorno scelto per le misurazioni dell'orientamento, il Sole produce immagini delle finestre in asse con le colonne. A quel momento il Sole aveva un azimut di 166 gradi, che corrisponde alla direzione del lato corto delle navate della chiesa.
Con un laser rosso a 656 nm, attraverso il foro si è verificato il tracciato in figura 3, verso l'azimut 121°, ricavato dalla mappa, una volta fissato l'orientamento del lato corto della chiesa con l'aiuto del Sole.[2]
Infatti il laser andava ad illuminare la quinta crociera della navata destra, in corrispondenza dell'attacco del muro destro della cappella, ed il suo percorso è riportato in figura 3 dalla linea lunga che parte dalla terza colonna di sinistra dove è situato il foro.

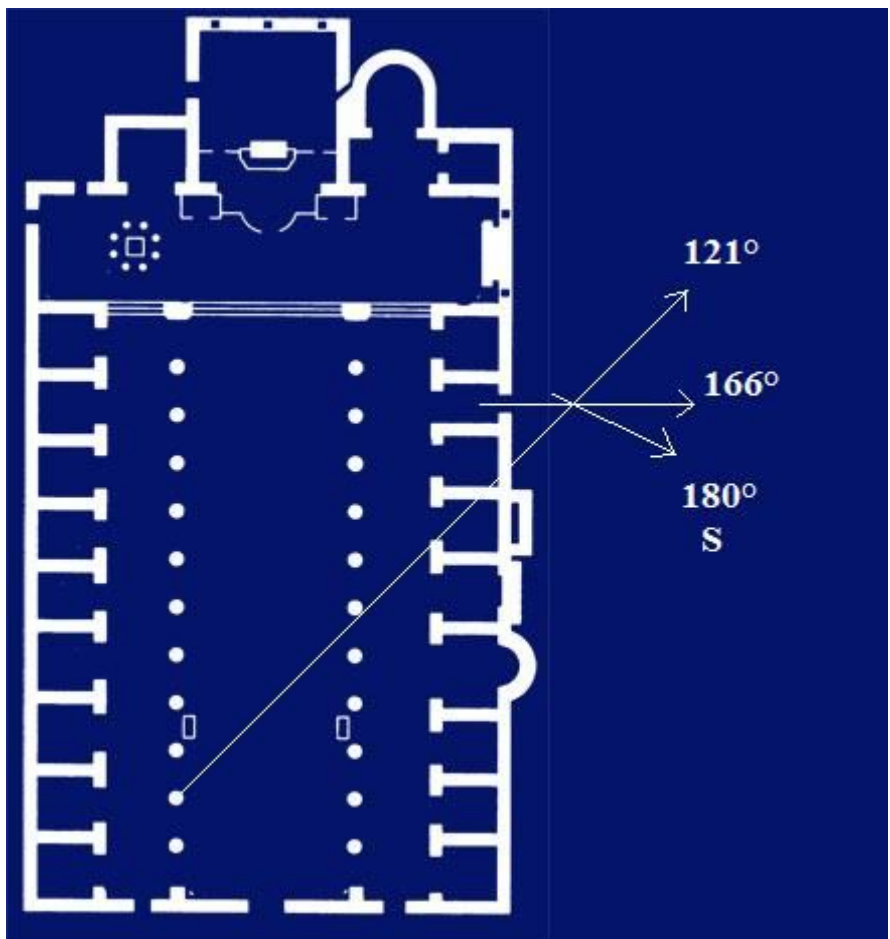

Fig. 4 gli azimut corrispondenti all'orientamentamento del lato corto della chiesa e del foro della terza colonna. E' riportato anche il Sud (180°).

**Le misure:**

La colonna in porfido ha un diametro di 243 cm all'altezza del foro.
Approssimandone la sezione ad un cilindro, il corrispondente diametro è di 77.4 cm.
L'altezza del foro è 1m 61 da un lato e 1m 83 dall'altro, con un dislivello totale di 22 cm; la lunghezza del foro è di 74 cm, corrispondente all'ipotenusa di un triangolo rettangolo il cui rimanente cateto è una corda di 70.65 cm del cerchio sezione del

---
[2] Il metodo per misurare l'allineamento di un edificio, sia interni che esterni, è ampiamente spiegato nel testo C. Sigismondi, Effemeridi, UPRA, Roma, 2008.

cilindro. In base alle misure appena descritte il foro non attraversa il centro della colonna.

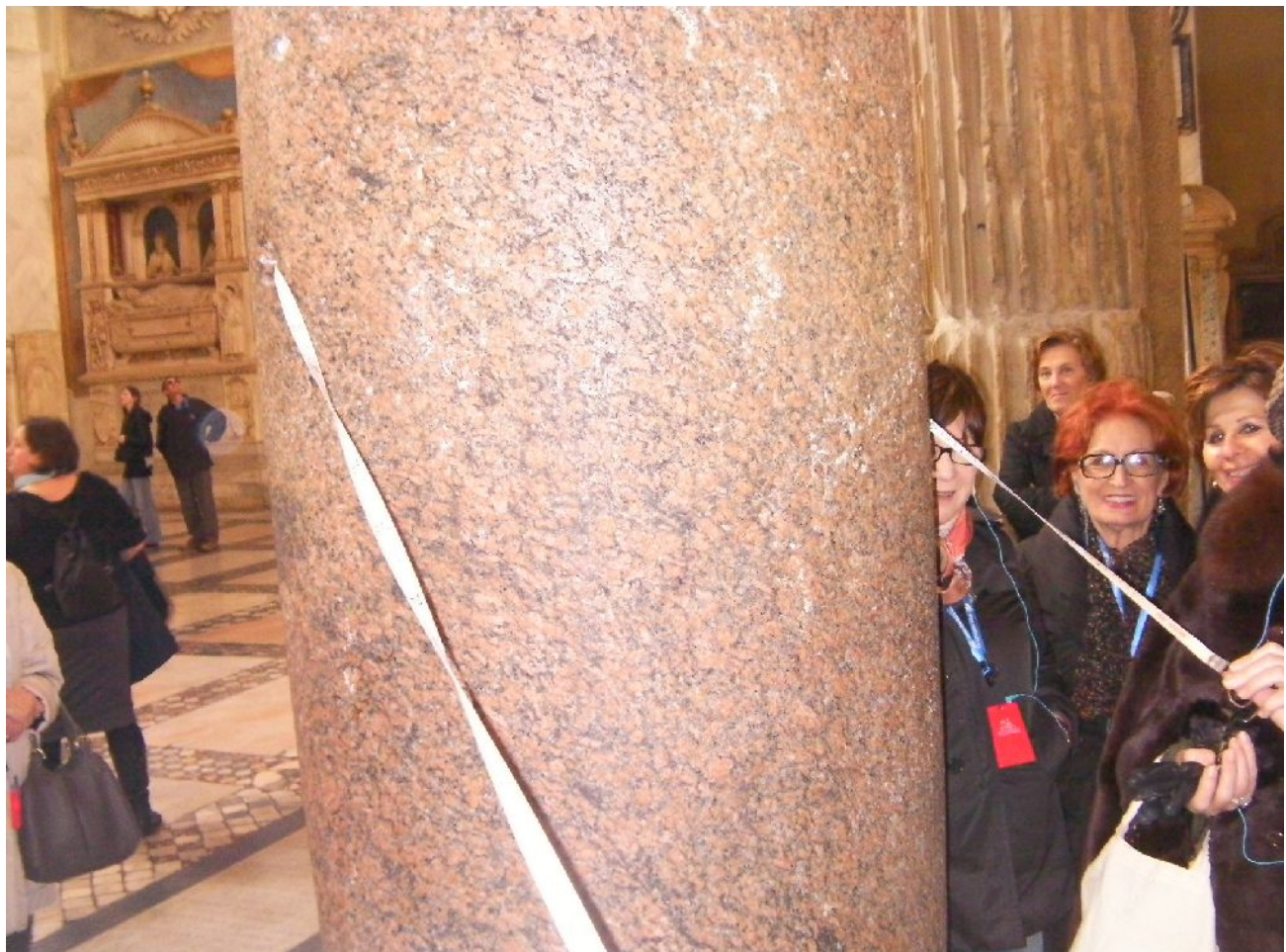

Fig. 5 La colonna durante le misure del 18 gennaio 2014.

Con questi lati per il triangolo rettangolo l'angolo che il tubo forato nella colonna forma con l'orizzonte risulta di 17.3°.

**Transito meridiano di un astro**

A Roma, latitudine 41.9° N se si pone in asse Nord-Sud quel tubo, viene puntata una regione di cielo corrispondente alla declinazione di 48.1°-17.3°=30.8° S.
Avendo il foro un diametro di 3 cm, il campo di vista è di 140'
Dunque se fosse puntato esattamente al Sud risulterebbero visibili stelle con declinazione δ=-30°48' ± 70'.
Il Sole al solstizio invernale raggiunge 23.5° S, quasi 24° al tempo di Augusto per effetto della variazione dell'obliquità dell'eclittica.
Dunque questo tubo, in meridiano, non avrebbe consentito di osservare il Sole.

Sappiamo che Nicola Cusano nel castello di Andraz realizzo' un foro di inclinazione 16.7° molto prossima a quella del foro della colonna dell'Aracoeli e puntato verso l'azimut 208.5°. Il De Dona' che ha studiato il foro di Cusano[3] ha formulato e provato l'ipotesi che il foro servisse per determinare, per interpolazione, la data del Solstizio di inverno, quando il Sole arrivava ad illuminare la parete dietro al foro, che fungeva da schermo. Nel lavoro sul foro di Cusano si cercano le stelle eventualmente visibili attraverso quel foro, ma non ne risulta nessuna brillante.

Nel caso dell'Aracoeli il campo di vista è più stretto e per effetto della latitudine il foro inquadra un settore di cielo più australe.

### Culminazione di un astro a 30° 48' Sud duemila anni fa

L'effetto della precessione[4] degli equinozi per un astro che duemila anni orsono fosse stato a declinazione 30°48' Sud dipende dalla sua ascensione retta, a seconda se si trova più o meno lontano dal polo sud dell'eclittica, dove lo spostamento è massimo.

Con l'aiuto di programmi quali Stellarium e Ephemvga è possibile osservare il cielo di 2000 anni fa ricercando stelle brillanti a declinazione 30° 48' S visibili da Roma.

La stella più brillante attorno a quella declinazione, duemila anni fa, è Shaula, Lamba Scorpii, di magnitudo 1.6 e declinazione 32°48S oppure Kaus Australis (epsilon Sgr) di 1.75 a 32°22 S o anche beta Ceti, 2.0, a 28°57 S. Eta Cen a 32°09 di 2.3, Adara 28°02 S di magnitudo 1.50.

Nessuna di queste stelle rientra nel campo di vista del tubo tra 29° 38' e 31° 58'.

Dunque anche l'ipotesi di transito meridiano stellare sembra sfumare.

### Passaggio del Sole ad azimut e altezza fissate

Resta la possibilità di passaggio del Sole, o di una stella, come nel caso del castello di Andraz, rispetto ad una direzione fissata.

A Roma l'altezza di 17.3° il Sole la raggiunge in qualsiasi giorno dell'anno, entro un range di azimut abbastanza ampio, che oggi va dai 90° ai 147° e dai 213° ai 270° e 2000 anni fa era leggermente più stretto 93°-144° e 216°-267° a causa della maggiore obliquità dell'asse terrestre.

L'azimut 116° con altezza 17.3° veniva raggiunto dal Sole il giorno 12 ottobre alle 8:00 ora solare nell'anno 1 dopo Cristo, sotto l'imperatore Ottaviano Augusto, con il Sole a declinazione -6°43'.

Oggi quell'azimut viene raggiunto il giorno 11 ottobre alla stessa ora, con il Sole a declinazione meridionale di -7°01'. Il 2 marzo si ripresenta alle 7:25 UT la situazione simmetrica rispetto al solstizio invernale.

Nessuna stella più brillante della magnitudine 3 ha una declinazione δ= -7°±70'. L'unica stella degna di nota che aveva questa declinazione 2000 anni fa è alfa Monocerotis, di magnitudo 3.9, che potrebbe essere osservata attraverso il foro nell'attuale orientamento.

---

3  http://www.icra.it/gerbertus/2010/Gerbertus_1-pp251-269-DeDona.pdf
4  http://www.archaeoastronomy.it/calcoloFK4.htm

# Conclusioni

Il tubo scavato nella colonna di porfido dell'Aracoeli merita l'attenzione degli appassionati e gli esperti di archeoastronomia, con maggiori informazioni rispetto alla singola riga che si trova nella «Guida rossa» del Touring Club Italiano.[5]

La stessa iscrizione sulla colonna porta un significato cruciale per la denominazione di tutta la basilica di S. Maria in Aracoeli, legandola alla tradizione del libro delle Sibille dell'apparizione della Vergine col Bambino ad Augusto e della conseguente costruzione dell' Aracoeli da parte dell'Imperatore; quindi lo studio del foro in quella stessa colonna acquisisce un senso davvero particolare.

Le misurazioni presentate in questo articolo contribuiscono a far luce sui possibili impieghi di questo foro in astronomia; escludendo il suo uso in meridiano sia con il Sole che con stelle brillanti.

Dato che la colonna non è più nel suo sito originale, è possibile immaginare altri usi, come quello ipotizzato per il foro nella stanza del Castello di Andraz attribuito a Nicola Cusano, tuttavia con l'incognita dell'azimut verso cui originariamente il foro era orientato delle soluzioni sono possibili per qualsiasi giorno dell'anno.

# Referenze

---

5 Nella terza colonna sinistra, iscrizione «a cubiculo Augustorum» e foro che trapassa diagonalmente la colonna, forse per misurazioni astronomiche. G. Morganti, in Guida d'Italia, Roma ed. TCI, Milano, 1993, p. 402.